\documentclass[twocolumn,showpacs,preprintnumbers,amsmath,amssymb]{revtex4}

\usepackage{graphicx}
\usepackage{dcolumn}
\usepackage{bm}
\usepackage[latin1]{inputenc}

\begin{document}

\title{Quantum dynamics of localized excitations in a symmetric trimer molecule}
\author{R. A. Pinto}
\author{S. Flach}
\affiliation{Max-Planck-Institut f\"ur Physik komplexer Systeme\\
N\"othnitzer Str. 38, 01187 Dresden, Germany }

\date{\today}

\begin{abstract}
\hspace{0.27cm}We study the time evolution
of localized (local bond) excitations
in a symmetric quantum trimer molecule.
We relate the dynamical 
properties of localized excitations such as their 
spectral intensity
and their temporal evolution (survival probability and tunneling of bosons)
to their degree of overlap with quantum tunneling pair states.
We report on the existence of degeneracy points in the trimer eigenvalue
spectrum for specific values of parameters due to avoided crossings
between tunneling pair states and additional states.
The tunneling of localized excitations which 
overlap with these degenerate states
is suppressed on all times. As a result local bond excitations may
be strongly localized forever, similar to their classical counterparts.

\end{abstract}


\maketitle

\section{Introduction}

The study of the classical and quantum dynamics of excitations in
non-linear systems with few degrees of freedom has been 
used for decades to understand the processes of energy redistribution
after an initial local bond excitation in polyatomic molecules
\cite{Uzer,Jaffe,keshavamurthy,Carney,Comp,Farantos}.
Equally, interest in these issues evolved from the more
mathematical perspective of nonlinear dynamics, localization
of energy and solitons \cite{EilbeckPhysicaD16}. 
This second path was boosted by the observation of
discrete breathers (DB) - time-periodic and spatially localized excitations -
in a huge variety of spatially discrete lattice systems
\cite{FlachPhysRep295,physicstoday,Sievers,AubryPhysicaD103}.
The flood of recent experimental observations of DBs in various systems
includes such different systems as bond excitations in molecules, 
lattice vibrations and spin excitations in solids, electronic currents in coupled
Josephson junctions, light propagation in interacting optical waveguides, 
cantilever vibrations in micromechanical
arrays, cold atom dynamics in Bose-Einstein condensates loaded on optical lattices,
among others \cite{SchwarzPRL83,SatoNature,SwansonPRL82,TriasPRL84,BinderPRL84,EisenbergPRL81,
FleischerNature422,SatoPRL90,TrombettoniPRL86,OstrovskayaPRL90,EiermannPRL92}.
In a substantial part of these cases quantum dynamics of excitations is either
unavoidable (molecules, solids) or reachable by corresponding
parameter tuning (Josephson junctions, Bose-Einstein condensates).

Progress in classical 
theory was achieved using a synergy of analytical results and
computational approaches. The computational aspect is vital because
we deal generically with non-integrable systems, which can not be completely
solved analytically. Computational studies of
classical systems of many interacting degrees of freedom (say $N$ oscillators) 
are straightforward since we have to integrate $2N$ coupled first-order
ordinary differential equations, so $N \sim 10^4$ is no obstacle to do even
long-time simulations in order to study statistical properties.
The quantum case is much less accessible by computational studies.
This is because in general each degree of freedom (e.g. an oscillator)
is now embedded in an
infinite-dimensional Hilbert space. Even after restricting to only $s$ states per oscillator,
the dimension of the Hilbert space of the interacting system is now $s^N$,
making it nearly impossible to treat both large values of $s$ and $N$ -
independent of whether we aim at integrating the time-dependent Schroedinger equation
or diagonalizing the corresponding Hamiltonian. 
However it is possible to treat small systems with $N=2,3$, which adds to the
above mentioned studies of bond excitations in molecules, perspective cases
of few coupled Josephson junctions, and Bose-Einstein condensates in optical
traps with just a few wells \cite{SmerziPRL79,RaghavanPRA59}.
Remarkably, in the last case there is already an experimental realization
\cite{AlbiezPRL95}.

Extensive studies of a dimer model $N=2$ with additional conservation
of the number of excitations (bosons) have been accomplished
\cite{ScottPhysLettA119,BernsteinPhysicaD68,Aubry,
BernsteinNonlin3,kalosakas1,kalosakas2,largespin1,largespin2}. The conservation of energy and boson number
makes this system integrable. Due to the nonlinearity of the model the invariance under
permutation of the two sites (bonds, spin flips etc)
is not preventing from having classical trajectories which are not invariant
under permutations. These trajectories correspond to a majority of bosons (and thus energy)
being concentrated on one of the sites.
Quantum mechanics reinforces the symmetry of
the eigenstates via dynamical tunneling in phase space (without obvious potential
energy barriers being present) \cite{Davis,Kesha2}. The tunneling time is inversely proportional
to the energy splitting of the corresponding tunneling pairs of eigenstates.
Notice that while most of the quantum computations concerned diagonalization of the
Hamiltonian, a few results show consistency with numerical integration of
the Schroedinger equation \cite{kalosakas2,Flach1}.
In \cite{AlbiezPRL95} the first experimental observation of
non-linearity-induced localization of Bose-Einstein condensates in a
double-well system was obtained, in agreement with results discussed above. 


The extension of the dimer to a trimer $N=3$ allows to study the 
fate of the tunneling pairs in the presence of nonintegrability  \cite{WrightPhysicaD69,
Flach1,Fillippo,Cruzeiro,Chefles,FlachPRB63}
and in an effective presence of the fluctuation of the number of bosons (on the dimer).
Diagonalization showed that tunneling pairs survive up to a critical strength 
of nonintegrability \cite{Flach1}, while the pair splittings showed 
characteristic resonances due to interactions with other eigenstates \cite{FlachPRB63}.

Trimer models have been also extensively studied in order to describe spectral
properties and then energy transfer in ABA molecules like water
\cite{Lawton,Child,Sibert,Schmid,Kellman}, which
is connected with the appearance of
quantum local modes (discrete
breathers). In these studies the presence of local modes was already
identified as nearly
degenerate eigenstates (tunneling pairs) in the 
eigenvalue spectra of the considered systems.  

In this work we study the time evolution of localized
excitations in the trimer, and compare with the spectral properties
of the system. We compute the eigenvalues and eigenstates of the quantum system and
then the expectation values of the number of bosons at
every site on the trimer and the survival probability of different initial
excitations as a function of time. We also compute the spectral intensity of
the initial excitations to see how many eigenstates overlap are involved.
That allows to draw conclusions about the correspondence between the
time evolution of a localized initial quantum state (not an eigenstate)
and the presence or absence of quantum breathers, i.e. dynamical tunneling eigenstates. 
We identify novel degeneracies in the trimer spectrum due to avoided crossings,
and relate these events to unusual classical-like behaviour of quantum localized excitations.

\section{Local bond excitations in the classical case}

The classical trimer is described by the Hamiltonian \cite{Flach1}
\begin{eqnarray}
H = H_d + \frac{1}{2}(P_3^2+X_3^2) + \frac{\delta}{2}(X_1X_3 + P_1P_3
{}\nonumber \\ + X_2X_3 + P_2P_3),
\end{eqnarray}
\begin{eqnarray}
H_d = \frac{1}{2}(P_1^2+P_2^2+X_1^2+X_2^2) + \frac{1}{8}[(P_1^2+X_1^2)^2 {}
\nonumber \\ +(P_2^2+X_2^2)^2] + \frac{C}{2}(X_1X_2+P_1P_2),
\end{eqnarray}
where $H_d$ is the dimer part. In all of this work we use dimensionless
quantities. $C$ is the coupling inside
the dimer, and $\delta$ is the coupling between site 3 and the dimer
which also destroys the integrability of the system. Using the transformation $\Psi_i = (1/\sqrt{2})(X_i+\mathrm{i}P_i)$ the Hamiltonian becomes
\begin{equation}
H = H_d + \Psi_3^* \Psi_3 + \delta( \Psi_1^*\Psi_3 + \Psi_2^*\Psi_3 + cc )
\end{equation}
\begin{eqnarray}
H_d =  \Psi_1^* \Psi_1 + \Psi_2^* \Psi_2 + \frac{1}{2}[
(\Psi_1^*\Psi_1)^2 + (\Psi_2^*\Psi_2)^2 ]  {}\nonumber \\
+ C (\Psi_1^*\Psi_2 +cc),
\end{eqnarray}
and the equations of motion transform to $\mathrm{i}\dot{\Psi}_i=\partial
H/\partial\Psi_i^*$.
Note that the total norm
$B = \Psi_1^* \Psi_1 + \Psi_2^* \Psi_2 + \Psi_3^* \Psi_3$
is conserved, and hence the problem is effectively two-dimensional. Also the
trimer (and the dimer) is invariant under permutation
of sites 1 and 2.

\begin{figure}[!t]
\includegraphics[width=3.4in]{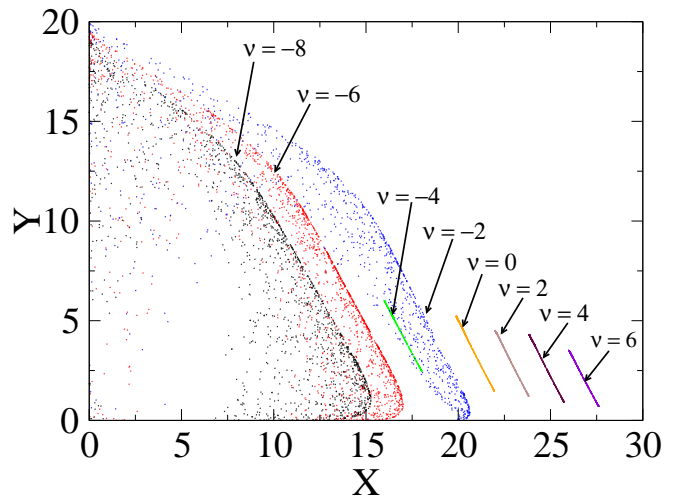}
\caption{\label{Poincare}(Color online) Poincare map of the classical phase space flow of the trimer. The
map condition is $\Delta_{13}=0$. The plotting plane is  $X=|\Psi_1|^2$ and
$Y=|\Psi_2|^2$. The parameters are $B=40,\, C=2,\, \delta=1$.}
\end{figure}

We are interested in the fate of localized excitations, where some energy is excited
e.g. on site 1, and none on site 2 (inside the dimer). The third site may have some nonzero
energy as well (like an environment).
For different initial conditions 
\begin{equation}
\Psi_1(0)=\sqrt{\frac{B}{2} + \nu},\;\;
\Psi_2(0)=0,\;\; \Psi_3(0)=\sqrt{\frac{B}{2} - \nu}
\end{equation}
we computed the time evolution of the quantities $|\Psi_i|^2=\Psi_i^*\Psi_i$
by numerically
solving the equations of motion. In
all computations we used $B=40,\,\,C=2$, and $\delta=1$.
We also generate a Poincare map 
(Fig.\ref{Poincare}) using the condition $\Delta_{13}=0$
($\Psi_i = A_ie^{\mathrm{i}\varphi_i},\,\Delta_{ij}=\varphi_i-\varphi_j$) and the plane
$X=|\Psi_1|^2,\,Y=|\Psi_2|^2$.
\begin{figure}[!h]
\includegraphics[width=3.4in]{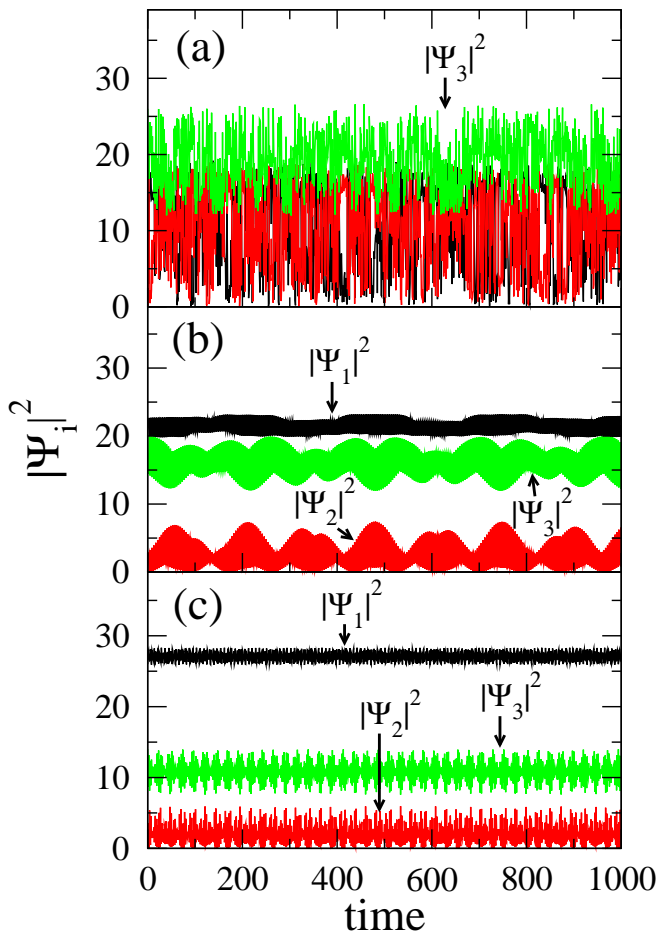}
\caption{\label{cNt1}(Color online) Time evolution of $|\Psi_i|^2$ ($i=1,2,3$) for different initial
states $\Psi_{1,3}(0)=\sqrt{20\pm\nu}$ ($\Psi_2(0)=0$):
(a) $\nu=-6$, (b) $\nu=0$, (c) $\nu=6$.}
\end{figure}
We observe that for positive $\nu$ the evolution is regular and not invariant
under permutation, so most of the energy initially placed on site 1 stays there,
with site 2 becoming only little excited.
Negative values of $\nu$ yield chaotic motion which is permutation invariant.

This transition from localization to delocalization of energy is also
nicely observed in the temporal evolution in Fig.\ref{cNt1}.
Increasing $\nu$ from negative to positive values the 
energy exchange between sites 1 and 2 of the dimer is stopped.

\section{Local bond excitations in the quantum trimer}

	The quantum trimer is obtained  after replacing the complex functions
$\Psi, \Psi^*$ by the bosonic operators $a$ and $a^\dagger$ (rewriting
$\Psi^*\Psi = (1/2)(\Psi^*\Psi+\Psi\Psi^*)$ previously to insure the invariance
under exchange $\Psi\leftrightarrow\Psi^*$):
\begin{equation}\label{eq:Ham}
\hat{H} = \hat{H}_d + \frac{3}{2} \hat{a}_3^{\dagger}\hat{a}_3 +  \delta( \hat{a}_1^{\dagger}\hat{a}_3 +
\hat{a}_2^{\dagger}\hat{a}_3 + c.c.),
\end{equation}
\begin{eqnarray}
\hat{H}_d = \frac{15}{8} + \frac{3}{2}(\hat{a}_1^{\dagger}\hat{a}_1 + \hat{a}_2^{\dagger}\hat{a}_2) +
\frac{1}{2}[(\hat{a}_1^{\dagger}\hat{a}_1)^2 +(\hat{a}_2^{\dagger}\hat{a}_2)^2] {} \nonumber\\ 
+ C(\hat{a}_1^{\dagger}\hat{a}_2 + c.c.),
\end{eqnarray}
where we take $\hbar=1$. The boson number operator $\hat{B}=\hat{a}_1^{\dagger}\hat{a}_1 +
\hat{a}_2^{\dagger}\hat{a}_2 + \hat{a}_3^{\dagger}\hat{a}_3$ commutes with
the Hamiltonian, so we may diagonalize (\ref{eq:Ham}) in the basis of
eigenfunctions of $\hat{B}$, $\{|n_1,n_2,n_3\rangle\}$, where $n_1,n_2,n_3$
respectively are the number of bosons at site
1, 2, and 3. There are $(b+1)(b+2)/2$ eigenstates in the subspace
corresponding to a fixed value of the eigenvalue $b$ of $\hat{B}$. Since the
Hamiltonian is invariant under permutation between sites 1
and 2 we expanded the wave function in the basis of symmetric and
antisymmetric eigenstates of $\hat{B}$,
$\{|n_1,n_2,n_3\rangle_{S,A}\}$, where
\begin{equation}
|n_1,n_2,n_3\rangle_{S,A}=\frac{1}{\sqrt{2}}(|n_1,n_2,n_3\rangle \pm |n_2,n_1,n_3\rangle).
\end{equation}
Then the initial state $|\Psi_0\rangle=|n_0,m_0,l_0\rangle$ writes as
\begin{eqnarray}
|\Psi_0\rangle &=& \frac{1}{\sqrt{2}}(|n_0,m_0,l_0\rangle_S +
|n_0,m_0,l_0\rangle_A),{}\nonumber\\
&\equiv& \frac{1}{\sqrt{2}}(|\Psi_0\rangle_S +
|\Psi_0\rangle_A),
\end{eqnarray}
In this representation diagonalization of the Hamiltonian
reduces to diagonalize two smaller matrices---symmetric and antisymmetric
decompositions of $\hat{H}$---whose eigenvalues are ${E_{\mu}^{(S,A)}}$, with
less computing cost than diagonalization of the full Hamiltonian. All
computations were done using this representation.

We computed the time evolution of expectation values of the number of bosons at every site on
the trimer $\langle n_i\rangle
(t)=\langle\Psi_t|\hat{n}_i|\Psi_t\rangle$ and the survival probability
$P_t=|\langle\Psi_0|\Psi_t\rangle|^2$ (see appendix for explicit expressions),
starting with various boson number
distributions among site 1 and site 3 controlled by the number $\nu$:
$|\Psi_0\rangle = |b/2+\nu, 0, b/2-\nu\rangle$, with
$b=40,\,C=2$, and $\delta=1$. In computations we dropped the two first terms
of the Hamiltonian, which are diagonal and just shift the spectrum.

\subsection*{Tunneling pairs and localization}

\begin{figure}
\includegraphics[width=3.1in]{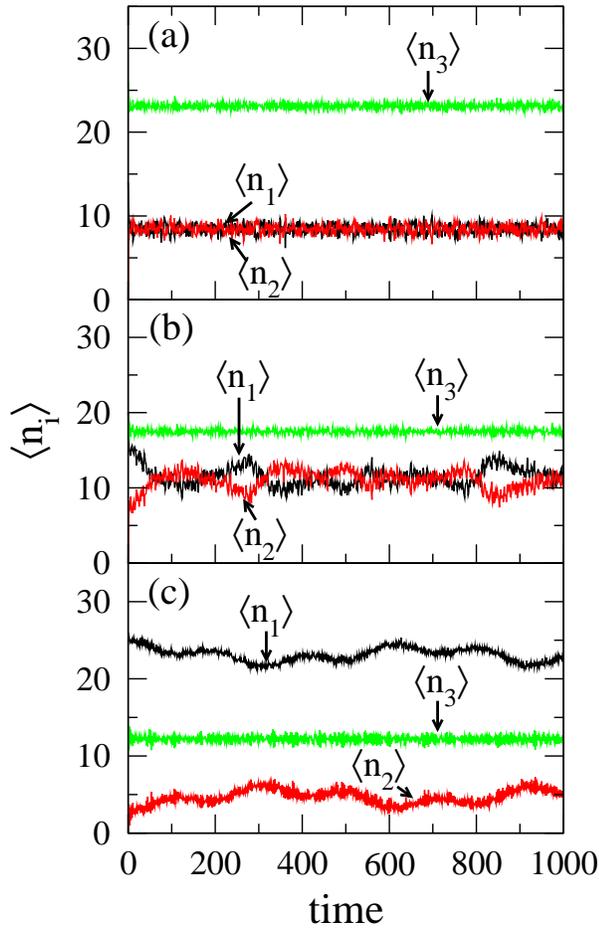}
\caption{\label{Nt1}(Color online) Time evolution of expectation values of the number of
bosons at each site of the
trimer for different initial states $|\Psi_0\rangle=|20+\nu,0,20-\nu\rangle$:
(a) $\nu=-6$, (b) $\nu=0$, (c) $\nu=6$.}
\end{figure}

\begin{figure}
\includegraphics[width=3.in]{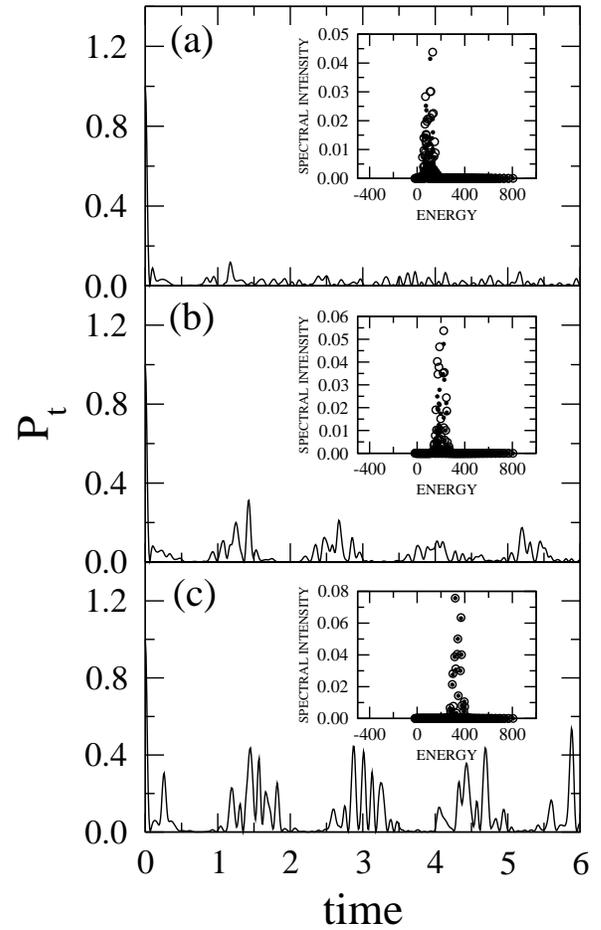}
\caption{\label{Pt}(Color online) Survival probability of the initial state $|\Psi_0\rangle =
|20+\nu,0,20-\nu\rangle$. (a) $\nu=-6$, (b) $\nu=0$, (c) $\nu=6$. Inset:
Spectral intensity of the initial state $|\Psi_0\rangle$. Filled
circles---symmetric eigenstates, open circles---antisymmetric eigenstates.}
\end{figure}

	In  Fig.\ref{Nt1} we show the time evolution of expectation values
of the number of bosons in the trimer. When the initial
excitation is mainly localized at the third site in the trimer there is a fast
redistribution of bosons between the two sites in the non-linear dimer until the 
dimer sites are
equally occupied ( Fig.\ref{Nt1}-a). As we place
more bosons on the dimer (site 1) the tunneling time of the excitation
increases rapidly until the time of computation becomes
too short to observe slow tunneling. On these timescales
we thus observe localization of bosons on one site in the dimer
( Fig.\ref{Nt1}-b and \ref{Nt1}-c), in analogy to the
classical case.
The reason for this behavior is the appearance
of {\it tunneling pairs} of symmetric and antisymmetric eigenstates with very close
eigenenergies in comparison to the mean energy separation between eigenstates
( Fig.\ref{crossing2}).
These pairs strongly overlap with the
initial state, as observable from the spectral intensity $I_{\mu}^0 =
|\phi_{\mu}^0|^2$ in the inset of the  Fig.\ref{Pt}.

	The results in  Fig.\ref{Pt} show an enhancement of the survival probability
with increasing boson number at site 1, which is consistent with the results
discussed above. The dominant tunneling pairs in the spectral intensity (inset
of  Fig.\ref{Pt}) give the main
contribution to the time dependence of the survival probability \cite{Flach1}.

\subsection*{Avoided crossings and degenerate eigenstates}

\begin{figure}
\includegraphics[width=2.9in]{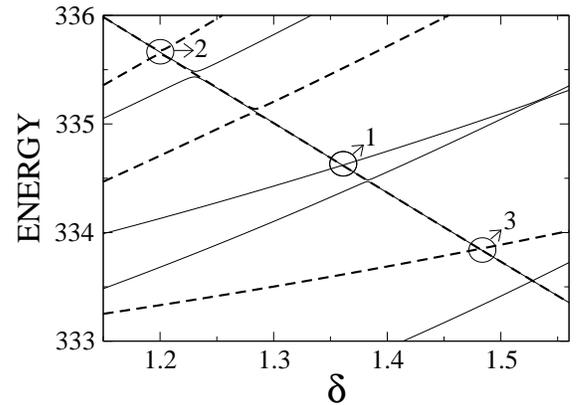}
\caption{\label{crossing2}(Color online) An enlargement of the trimer spectrum showing three particular pair-single
state interactions (numbered circles). Thin solid
line---symmetric eigenstates; thick dashed lines---antisymmetric
eigenstates.}
\end{figure} 
	Energy levels exhibit avoided crossings when we vary the parameter
$\delta$ which
regulates the strength of nonintegrability of the system \cite{Flach1}, as shown in
the  Fig.\ref{crossing2}.
Of particular interest is the outcome of the interaction of a single eigenstate
and a tunneling pair. 
The principal difference between these states is that the member of a tunneling
pair has exponentially small weight in the dynamical barrier region,
which is roughly defined by $n_1=n_2$ in the $n_1-n_2$ plane. A single
eigenstate will in general have much larger weight in this region.
Since each eigenstate is either symmetric or antisymmetric,
and a tunneling pair consists always of states with both symmetries, the interaction
with a third eigenstate will in general allow for an exact degeneracy
of two states with different symmetries. While this is in principle
possible for any two states of different symmetry, the exponentially small weight
of the tunneling pair states in the dynamical barrier region makes a difference.
Indeed, a linear combination of two states with large (not exponentially small)
weight in a barrier region yields again, though an assymetric state, but one
with large weight in the barrier region. Contrary, for the case
of a tunneling pair and a single state, we may expect
an asymmetric eigenstate which has much less weight in the barrier region,
leading to a much stronger localization of the state similar to a classical
one.

\begin{figure}
\includegraphics[width=3.4in]{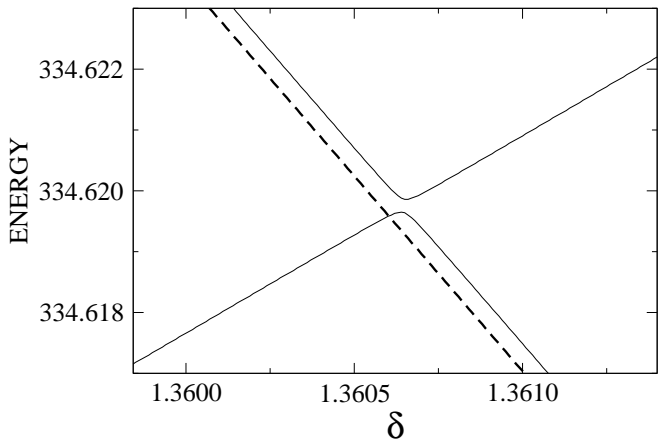}
\caption{\label{DEA268}(Color online) An enlargement of the trimer spectrum around the
avoided crossing 1 in figure \ref{crossing2}, involving the antisymmetric
state A-268 (thick dashed line) and the
symmetric states S-286 and S-285 (thin solid lines).}
\end{figure}

\begin{figure}
\includegraphics[width=3.4in]{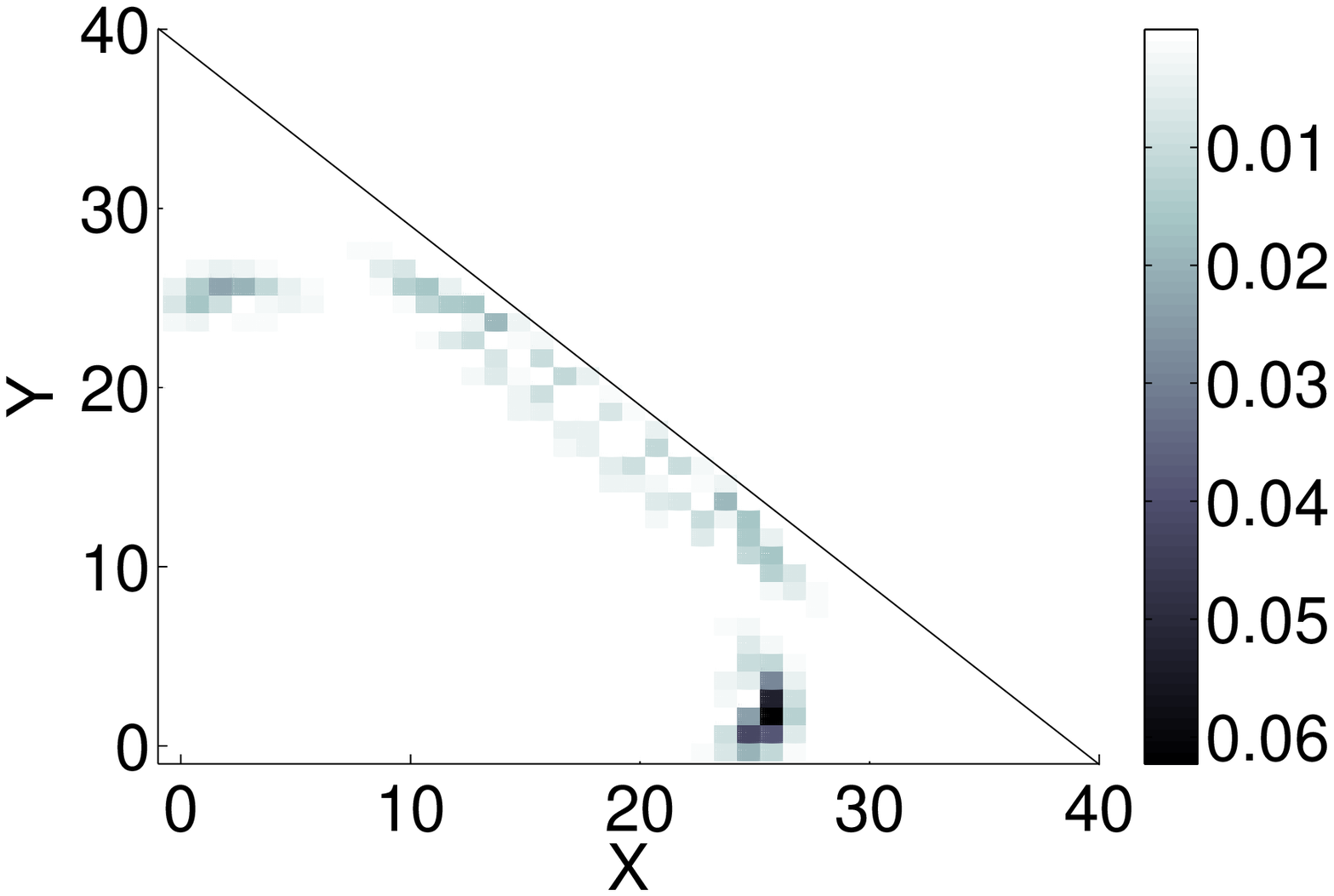}
\caption{\label{color}(Color online) Density of the asymmetric state
$|\phi\rangle=(|\phi_{285}^{(S)}\rangle + |\phi_{268}^{(A)}\rangle)/\sqrt{2}$  as a function of
the number of bosons at sites 1 and 2 at the degeneracy point $\delta_d$. Here
X is the number of bosons at site 1, and Y is the number of bosons at site 2.}
\end{figure}

We analyze three particular avoided crossings identified by numbered circles
in Fig.\ref{crossing2} by computing the energy
separation $\Delta E(\delta)$ between such a single state and a quantum breather tunneling pair. We
identify three different situations.
The first one shows that the energy levels intersect once in some degeneracy point 
(Fig.\ref{DEA268}).
At some value of the parameter $\delta$ the energy separation between one of
the members of the tunneling pair and the single state vanishes.
Tunneling is suppressed completely, and then an asymmetric linear
combination of the degenerate eigenstates will constitute a non-decaying
localized state. This situation has been well described by perturbation theory 
\cite{FlachPRB63}, where
effects of other eigenstates have been neglected.
We computed the density
$\rho(n_1,n_2) = |\langle n_1,n_2,n_3|\phi\rangle|^2$ of the asymmetric
eigenstate $|\phi\rangle=(|\phi_{d}^{(S)}\rangle +
|\phi_{d}^{(A)}\rangle)/\sqrt{2}$, where $|\phi_{d}^{(S)}\rangle$ and
$|\phi_{d}^{(A)}\rangle$ are the degenerate eigenstates. The result is shown in
the  Fig.\ref{color} where we can see that there is only a partial localization
of the excitation, since the wave function has visible contributions around the
diagonal $X=Y$ ($n_1=n_2$). Note that in addition it also shows sizable
contribution on the other side of the barrier ($n_1\approx 2,\; n_2\approx 26$
in Fig.\ref{color}). In fact the expectation values of the number of bosons for this state are
$\langle n_1\rangle = 14.99,\; \langle n_2\rangle = 14.89,\; \langle
n_3\rangle = 10.12$. Thus in terms of averages practically no localization
occurs since $\langle n_1\rangle \approx \langle n_2\rangle$ despite the
observable asymmetry in Fig.\ref{color}.
\begin{figure}
\includegraphics[width=3.4in]{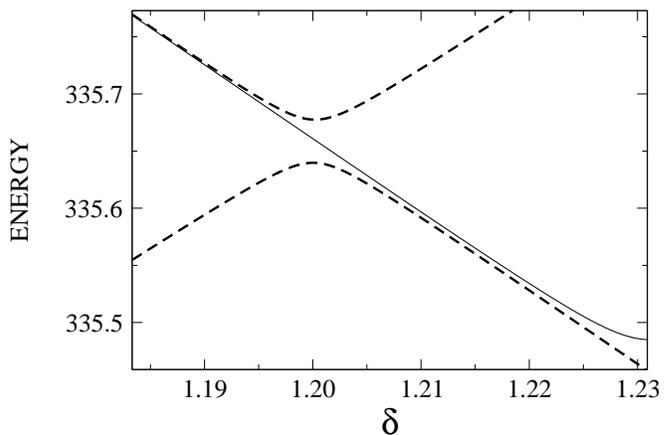}
\caption{\label{DES287}(Color online) An enlargement of the trimer spectrum around the
avoided crossing 2 in figure \ref{crossing2}, involving the symmetric
state S-287 (thin solid line) and the
antisymmetric states A-269 and A-270 (thick dashed lines).}
\end{figure} 
\begin{figure}
\includegraphics[width=3.4in]{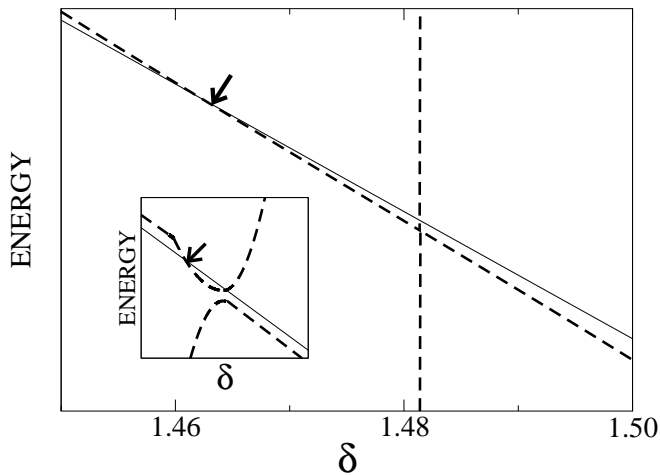}
\caption{\label{DES284}(Color online) An enlargement of the trimer spectrum around the
avoided crossing 3 in figure \ref{crossing2}, involving the symmetric
state S-284 (thin solid line) and the
antisymmetric states A-267 and A-268 (thick dashed lines). The curves for antisymmetric eigenstates
were generated from the data $\Delta E(\delta)$ by using
$E_A=E_S - \gamma \Delta E$, with $\gamma=1000$, for better visualization of
intersection between eigenvalues. Inset: Sketch
of the variation of eigenvalues
participating in the avoided crossing. Thin solid line---symmetric eigenstate,
thick dashed line---antisymmetric eigenstates. The arrows mark the analyzed
degeneracy point (see text).}
\end{figure} 
\begin{figure}
\includegraphics[width=3.4in]{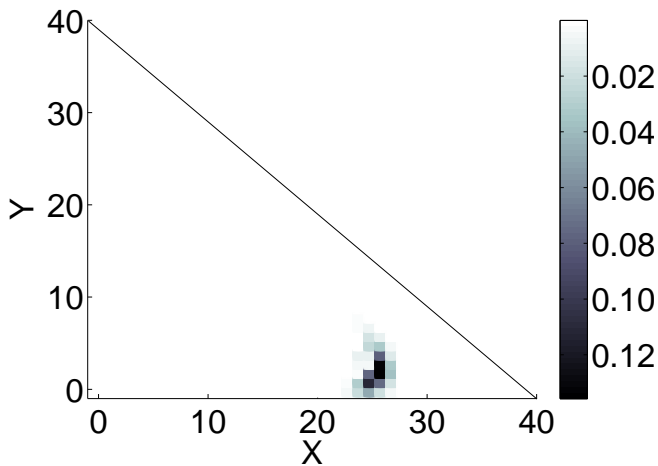}
\caption{\label{SA284}(Color online) Density of the asymmetric state
$|\phi\rangle=(|\phi_{284}^{(S)}\rangle + |\phi_{268}^{(A)}\rangle)/\sqrt{2}$  as a function of
the number of bosons at sites 1 and 2 at the degeneracy point $\delta_d$. Here
X is the number of bosons at site 1, and Y is the number of bosons at site 2.}
\end{figure} 
 
The other two cases appear as a consequence of the influence of
other states in the spectrum. In one case the energy levels do not
intersect at all ( Fig.\ref{DES287}), due to the presence of another avoided crossing
located nearby.
In the third case surprisingly we observe that the energy levels intersect
twice. The situation is shown in Fig.\ref{DES284}, and is sketched in the inset of the figure.
Due to the interaction with other states of the system we observe an intersection
of the two states of the tunneling pair at some distance from the actual
avoided crossing with the third state. Consequently both states have
exponentially small weight in the barrier region, and we may expect
a very strong localization. 
In Fig.\ref{SA284}
we can see that the asymmetric quantum breather in the degeneracy point
$\delta\simeq 1.462$ in Fig.\ref{DES284} (see arrow) is strongly localized and the
tunneling is suppressed for all times. Note that in both cases two and three
the order of
the participating levels before and after the avoided
crossing is not conserved, at variance to the first case we
discussed above and which was described also in reference \cite{FlachPRB63}.
The abovementioned strong localization of this exact asymmetric eigenstate
is reflected in the fact that the wave function has practically zero weight
around the barrier region $X=Y$ ($n_1=n_2$). Note that at variance with
Fig.\ref{color}, here the wave function has no sizable contribution on the
other side of the barrier as well. This state is thus very close to its
classical discrete breather counterpart (Fig.\ref{Poincare}). Indeed, for this
state $\langle n_1\rangle = 25.62,\; \langle n_2\rangle = 2.38,\; \langle
n_3\rangle = 12.00$. Consequently we find a very strong localization for the
expectation values, in addition to the observed asymmetry in Fig.\ref{SA284}.


\begin{figure}
\includegraphics[width=3in]{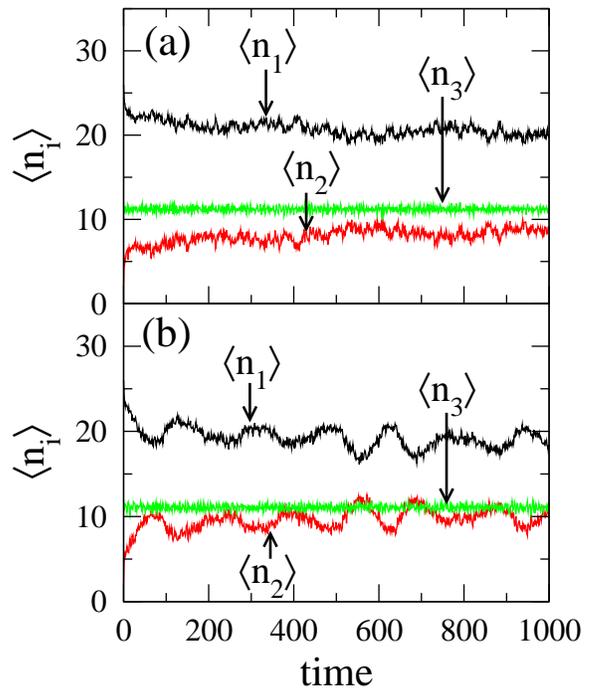}
\caption{\label{Ntdegen}(Color online) Time evolution of expectation values of the number of
bosons at every site on the trimer at the degeneracy points for the cases
shown in (a)  Fig.\ref{color}, (b)  Fig.\ref{SA284}. The initial state is
$|\Psi_0\rangle = |26,2,12\rangle$.}
\end{figure} 
\begin{figure}
\includegraphics[width=2.9in]{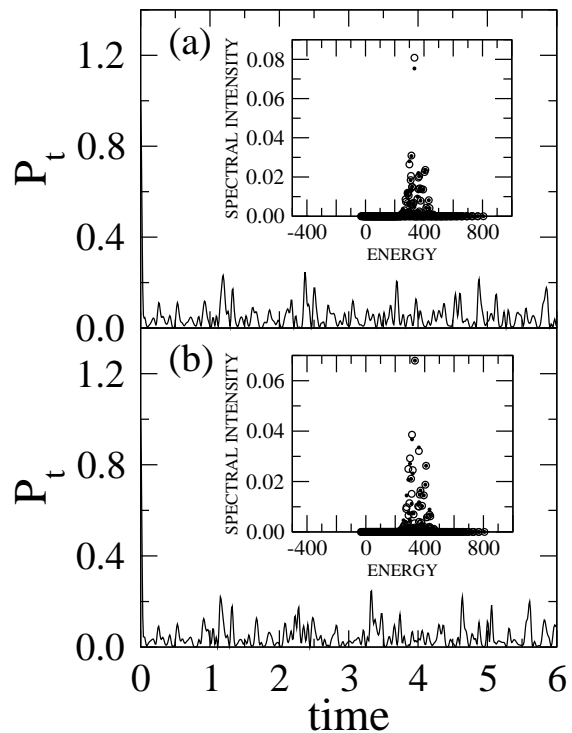}
\caption{\label{Ptdegen}(Color online) Survival probability of the initial state
$|\Psi_0\rangle = |26,2,12\rangle$ at the degeneracy points for the cases shown
in (a)  Fig.\ref{color}, (b)  Fig.\ref{SA284}. Inset: Spectral intensity
of the initial state $|\Psi_0\rangle$. Filled circles---symmetric eigenstates,
open circles---antisymmetric eigenstates.}
\end{figure} 

It is interesting to test whether
initial states with some distribution of bosons at every site
of the trimer ($|\Psi_0\rangle=|n_0,m_0,l_0\rangle$) can significantly
overlap with the above described asymmetric eigenstates.  
This distribution is given by the maxima in the density for every
asymmetric eigenstate (around $n_1=26$, and $n_2=2$). In figures \ref{Ntdegen}
and \ref{Ptdegen} we show the time evolution of the expectation value of the
number of bosons at every site on the trimer and the survival probability of such
an initial excitation. A
detailed analysis of the spectral intensity of the initial state
$|\Psi_0\rangle = |26,2,12\rangle$ (inset in the  Fig.\ref{Ptdegen}) shows
that this initial excitation overlaps strongly with the degenerate
eigenstates corresponding to the strong localization shown in Fig.\ref{SA284}. 
It implies that this excitation ( Fig.\ref{Ntdegen}-b) will
never distribute its quanta evenly over both sites of the
dimer. For the case shown in  Fig.\ref{Ntdegen}-a the initial
excitation has a smaller overlap with the degenerate eigenstates
which gives the partial localization shown in  Fig.\ref{color}. 
Since the overlap is not zero the excitation will also stay localized in the
sense that the crossing of curves corresponding to $\langle
n_{1} \rangle$ and $\langle n_{2} \rangle$ as in  Fig.\ref{Nt1}-a and
\ref{Nt1}-b will never occur. Note that despite the difference between
the analyzed cases one and three, the evolution of the expectation values
and the survival probabilities do not differ drastically. It needs more
sensitive details in the preparation of an initial state to observe
a practically total localization of bosons on one of the dimer sites
for case three as compared to case one. 

\section{Conclusions}

In this work we observed how spectral
properties of the Hamiltonian are reflected in the
time evolution of different localized excitations in a trimer molecule model by monitoring
the spectrum, the time evolution of expectation values of the number of bosons at every site on
the trimer and survival probabilities of different localized excitations.

The tunneling pair splitting determines the lifetime of
localized excitations. The survival probability and the time evolution of the expectation
values of the number of bosons are clear indicators for a localized excitation
being close or far from a quantum breather tunneling pair, while the
spectral intensity of localized excitations is typically broad and does not
show the peculiarities of the tunneling dynamics.
Probing the time evolution of initially localized excited states
thus allows to conclude about the presence or absence of tunneling pair
eigenstates.

We report on the
existence of degenerate levels in the spectrum due to the presence of
both avoided crossings and tunneling pairs.
In these degenerate points tunneling is
suppressed for all times. While in general the asymmetric
exact eigenstates will have quite large weight in the dynamical
barrier region (contributed by the single level), we observe
specific parameter cases where the weight is very small
and the corresponding asymmetric eigenstate very strongly localized.
Full or partial localization of bosons
appears for all time scales for some specific states and some
specific values of the parameters.
This effect could be studied in experimental situations of Bose-Einstein condensates
in few traps which weakly interact, as well as in systems of few coupled
Josephson junctions which operate in the quantum regime. Tuning experimental
control parameters will allow to lock localized excitations for specific
values and both prevent the excitation from tunneling, as well as allowing
for a fine tuning of the tunneling frequency from a small value down to zero
in the vicinity of these specific control parameter values.
\acknowledgments
We thank G. Kalosakas, S. Keshavamurthy, M. Johansson, and L. Schulman for useful discussions.
\\

\appendix*

\section{Expectation values and survival probability}

	Expanding the wave function in the basis of symmetric and
antisymmetric eigenstates of the Hamiltonian

\begin{eqnarray}
|\Psi_t\rangle = \sum_{\mu}\phi_0^{\mu (S)}e^{-iE_{\mu}^{(S)}t}|\phi^{\mu
(S)}\rangle + {} \nonumber \\
\sum_{\nu}\phi_0^{\nu (A)}e^{-iE_{\nu}^{(A)}t}|\phi^{\nu
(A)}\rangle ,
\end{eqnarray}
where $\phi_0^{\mu (S,A)}=\langle\phi^{\mu
(S,A)}|\Psi_0\rangle_{S,A}$ and $\phi_{n_1,n_2,n_3}^{\mu (S,A)} =
\langle\phi^{\mu}|n_1,n_2,n_3\rangle_{S,A}$,
the expectation value of the number of bosons at site $i$ writes as
\begin{equation}
\langle n_i\rangle (t)= \langle n_i^{(S)}\rangle (t) +
\langle n_i^{(A)}\rangle (t) + \langle n_i^{(M)}\rangle (t),
\end{equation}
where
\begin{eqnarray}
\langle n_1^{(S,A)}\rangle (t)&=& \frac{1}{4}\sum_{\mu,\mu^{\prime}}
\phi_{0}^{\mu (S,A)} \bar{\phi}_0^{\mu^{\prime}(S,A)}
e^{i(E_{\mu}^{(S,A)}-E_{\mu^{\prime}}^{(S,A)})t} {}\nonumber\\
&&\;\;\;\;\;\;\;\;\;\;\;\;\;\;\; \times F_{\mu,\mu^{\prime}}^{(S,A)}, {}\\
\nonumber \\
\langle n_2^{(S,A)}\rangle (t) &=& \langle n_1^{(S,A)}\rangle (t), {}\\
\nonumber \\
\nonumber\\
F_{\mu,\mu^{\prime}}^{(S,A)}\, &=& \sum_{ \{n_i\}_{S,A}}
\bar{\phi}_{n_1,n_2,n_3}^{\mu (S,A)}(n_1+n_2)
\,\phi_{n_1,n_2,n_3}^{\mu^{\prime}(S,A)}, {}\\
\nonumber\\
\nonumber\\
\langle n_1^{(M)}\rangle (t) &=& \Re \Big\{\frac{1}{2}\sum_{\mu,\nu}
\phi_{0}^{\mu (S)} \bar{\phi}_0^{\nu (A)}
e^{i(E_{\mu}^{(S)}-E_{\nu}^{(A)})t} {} \nonumber\\
&&\;\;\;\;\;\;\;\;\;\;\;\;\;\;\; \times F_{\mu,\nu}^{(M)} \Big\},{} \\
\nonumber \\
\langle n_2^{(M)}\rangle (t) &=& -\langle n_1^{(M)}\rangle (t),{} \\
\nonumber \\
\nonumber\\
F_{\mu,\nu}^{(M)}\, &=&  \sum_{\{n_i\}_A}
\bar{\phi}_{n_1,n_2,n_3}^{\mu (S)}(n_1-n_2) \,\phi_{n_1,n_2,n_3}^{\nu (A)},
\end{eqnarray}
\begin{eqnarray}
\langle n_3^{(S,A)}\rangle (t)&=& \frac{1}{2}\sum_{\mu,\mu^{\prime}}
\phi_{0}^{\mu (S,A)} \bar{\phi}_0^{\mu^{\prime}(S,A)}
e^{i(E_{\mu}^{(S,A)}-E_{\mu^{\prime}}^{(S,A)})t} {} \nonumber\\
&&\;\;\;\;\;\;\;\;\;\;\;\;\;\;\; \times G_{\mu,\mu^{\prime}}^{(S,A)}, {}\\
\nonumber \\
\nonumber\\
G_{\mu,\mu^{\prime}}^{(S,A)}\, &=& \sum_{\{n_i\}_{S,A}}
\bar{\phi}_{n_1,n_2,n_3}^{\mu (S,A)}(n_1+n_2) {} \nonumber\\
&&\;\;\;\;\;\;\;\;\;\;\;\;\;\;\;
\times\phi_{n_1,n_2,n_3}^{\mu^{\prime}(S,A)}, {}\\
\nonumber\\
\langle n_3^{(M)}\rangle (t) &=& 0,
\end{eqnarray}
where bars mean complex conjugation.

\end{document}